\documentclass[]{emulateapj}

\renewcommand{\tabcolsep}{3pt}

\slugcomment{To appear in {\it The Astrophysical Journal Letters.}}

\lefthead{Wilson et al.}
\righthead{The Cavity of Cygnus A}

\begin{document}

\title{The Cavity of Cygnus A}

\author{A. S. Wilson\altaffilmark{1, 2},
David A. Smith\altaffilmark{1},
 and
A. J. Young\altaffilmark{3}
}

\altaffiltext{1}{Astronomy Department, University of Maryland, College Park,
MD 20742; e-mail wilson@astro.umd.edu; dasmith@astro.umd.edu}

\altaffiltext{2}{Adjunct Astronomer, Space Telescope Science Institute, 3700
San Martin Drive, Baltimore, MD 21218}

\altaffiltext{3}{MIT Kavli Institute for Astrophysics and Space Research, 77 Massachusetts Avenue,
Cambridge, MA 02139; e-mail ayoung@space.mit.edu}
\begin{abstract}

In this paper, we focus on the limb-brightened, prolate spheroidal cavity of
the radio galaxy Cygnus A, as revealed by the Chandra X-ray Observatory.
We use the shock heated, thermal intracluster medium around the expanding
cavity to infer the properties of the radio synchrotron-emitting
gas inside the cavity.
The
gas along the N and S edges of the cavity is found to have an average 
temperature of 6.0 keV, which is hotter than the temperature (4.6 keV) of the
adjacent intracluster gas. It is proposed that this hotter gas is intracluster 
gas shocked by the expanding cavity. The shock is thus inferred to be weak
(Mach number 1.3, a value also inferred from the density jump at the cavity
edge) and its velocity 1,460 km s$^{-1}$. The total
kinetic power of the expansion is found to be 1.2 $\times$ 10$^{46}$ erg
s$^{-1}$, which is larger than both the total radio power and the
power emitted by the entire intracluster medium in the 2 -- 10 keV band. It
appears that most of the power of the jets in Cygnus A is currently going into
heating the intracluster medium. From the derived pressure inside the cavity,
there is no conclusive evidence for a component contributing pressure
additional to the magnetic fields and relativistic particles responsible
for the synchrotron radio emission.  Further, the ratio of energy densities in
positive to negative cosmic rays in Cygnus A is between 1 and 100 (the value in 
our Galaxy).

\end{abstract}


\section{Introduction}

It is now generally accepted that the lobes of powerful radio galaxies are
fuelled by two oppositely directed relativistic beams emitted from the
nucleus. By comparing the power supplied by the beams with the power
radiated by the radio components, Scheuer (1974) showed that only a small
fraction of the power supplied can be radiated away from the tip of the beam,
and that most of the energy must be deposited in a cavity surrounding the
beam. The dynamics of this cavity (or ``cocoon'' as it is sometimes called)
and its interaction with the external medium have been extensively discussed
(e.g. Scheuer 1974; Begelman \& Cioffi 1989; Kaiser \& Alexander 1997, 1999;
Heinz, Reynolds \& Begelman 1998; Alexander 2000; Reynolds, Heinz \&
Begelman 2001).

In order to derive basic properties of the cavity, such as its pressure and
expansion velocity, previous workers have had to use the synchrotron radio
emission, often assuming equipartition of energy between magnetic fields and
relativistic particles. There is no guarantee that this is the case and there
may be other sources of pressure in the cavity, such as low energy cosmic rays 
or diffuse hot gas. A more direct probe of the properties of the cavity could
be achieved if thermal emission from the shock driven by the overpressured
cavity into the surrounding gas could be observed.

In this paper, we deduce the presence and properties of the cavity and its
external (bow) shock from Chandra X-ray observations of the thermal emission
associated with the powerful radio galaxy Cygnus A. Deficits of X-ray emission
coincident with the inner radio lobes of Cyg A were first found by Carilli,
Perley \& Harris (1994) from ROSAT observations. However, ROSAT did not
have the spatial or spectral resolution to derive the detailed morphology or
gas temperatures in the vicinity of the cavity. The presence of a bow shock
ahead of radio hot spot B of Cyg A was inferred by Carilli, Perley \& Dreher
(1988) from measurements of Faraday rotation.

In the present paper, Section 2 reviews the Chandra observations, while
Section 3 derives various properties of the expanding cavity and the shock
driven into the intracluster medium. Section 4 summarizes our conclusions.

\section{Chandra Observations}

\begin{figure}
\epsscale{1.0}
\plotone{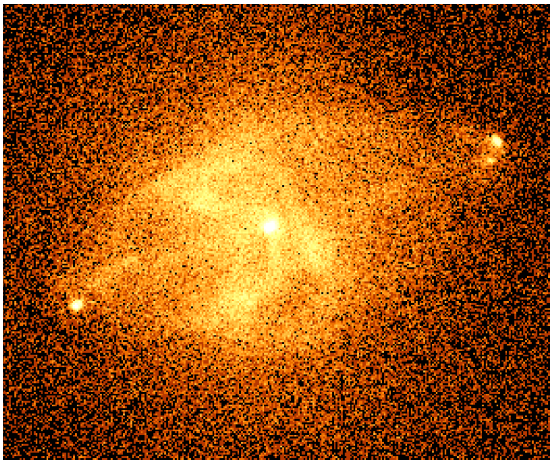}
\caption{The Chandra image on the scale of the radio source. The width of the
image is 2\farcm5.}
\label{Fig. 1}
\end{figure}

The Chandra observations of Cyg A have been described in our earlier papers
(Wilson et al. 2000; Young et al. 2002; Smith et al. 2002, hereafter Paper III).
For consistency with most recent discussions of the topic of this paper,
we use H$_{\rm 0}$ = 75 km s$^{-1}$ Mpc$^{-1}$ and q$_{\rm 0}$ = 0
(unlike Papers I -- III which assumed H$_{\rm 0}$ = 50 km s$^{-1}$ Mpc$^{-1}$
and q$_{\rm 0}$ = 0). The luminosity distance to Cyg A is then d$_{\rm L}$ =
231 Mpc and the angular size distance d$_{\rm A}$ = 207 Mpc, so 
1$^{\prime\prime}$ corresponds to 1.0 kpc. Preliminary remarks on Chandra
observations of the cavity have been presented in conference
proceedings (Wilson et al. 2002, 2003).

For illustration purposes, Fig. 1 shows the Chandra image of Cyg A. The
elliptical shaped morphology is interpreted in terms of hot gas swept up by the
cavity. Fig. 2 shows 6 cm VLA radio contours superposed on the X-ray grey scale.
It can be seen that the radio lobes are associated with relatively faint X-ray
emission and project inside the limb-brightened edge of the cavity, both
of which are in
accord with expectation.

\begin{figure}
\includegraphics[angle=270,scale=0.32]{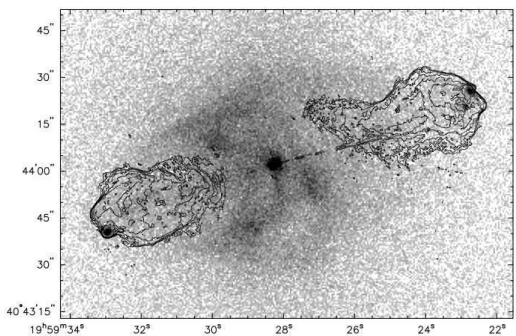}
\caption{Contours of a VLA 6 cm radio map (Perley et al. 1984) superposed on a
grey scale image of the X-ray emission.}
\label{Fig. 2}
\end{figure}

\section{Expansion of the Cavity}

In Paper III, we deprojected the X-ray spectra of the intracluster gas around
Cygnus A taken from 12 elliptical and circular annuli outside the cavity in 
order to derive the run of temperature, metallicity, density and pressure as a
function of radius. The goal was to derive the ambient values of the parameters 
presumed unaffected by the energy input by Cygnus A itself. The temperature of 
the X-ray emitting gas was found to drop from $\simeq$ 8 keV more than
$\simeq$ 80 kpc
from the center to $\simeq$ 5 keV some 50 kpc from the center, with the coolest
gas immediately outside the cavity. Since this work was published, improved 
response functions (``fefs'') have become available. We have, therefore,
deprojected each cluster shell and modelled
each one with a single temperature MEKAL plus foreground emission, absorbed
by a column of cold gas, precisely as done in Paper III but using the new fefs.
Such a model provides an acceptable fit to the spectra of all 12 shells, and
gives small changes in the run of parameters with radius compared with the
results presented in Paper III.
In particular, the (deprojected)
temperature in the innermost shell, just outside the cavity, is now
$T_{\rm 1}$ = 4.60 $\pm$ 0.5 keV, consistent with the value of 4.91
$\pm$ 0.6 keV given in Table 4 of paper III.

Fig. 3 shows the cavity with areas emitting bright X-ray emission outlined.
The updated temperatures of MEKAL models are indicated. These regions are so
bright that projection of foreground and background cluster emission onto them
is negligible. In Fig. 3, regions
with temperature $>$ 5 keV are outlined, and the temperature labelled, in blue,
while red is used for regions with temperatures $<$ 5 keV. It is immediately
noteable that the four regions projecting along the N and S edges of the 
cavity are hot (5.2 -- 6.8 keV), while those in the ``belts'' crossing the
cavity and the nucleus of the galaxy from NE to SW are cool (3.8 -- 4.4 keV).
The ``belts'' (which are not discussed in this paper) appear to be a disklike 
structure of cooling gas, seen obliquely and being accreted into the galaxy.
This cooling gas may form the outer part of the accretion disk which fuels the
nuclear activity.

\begin{figure*}
\includegraphics[angle=270,scale=0.6]{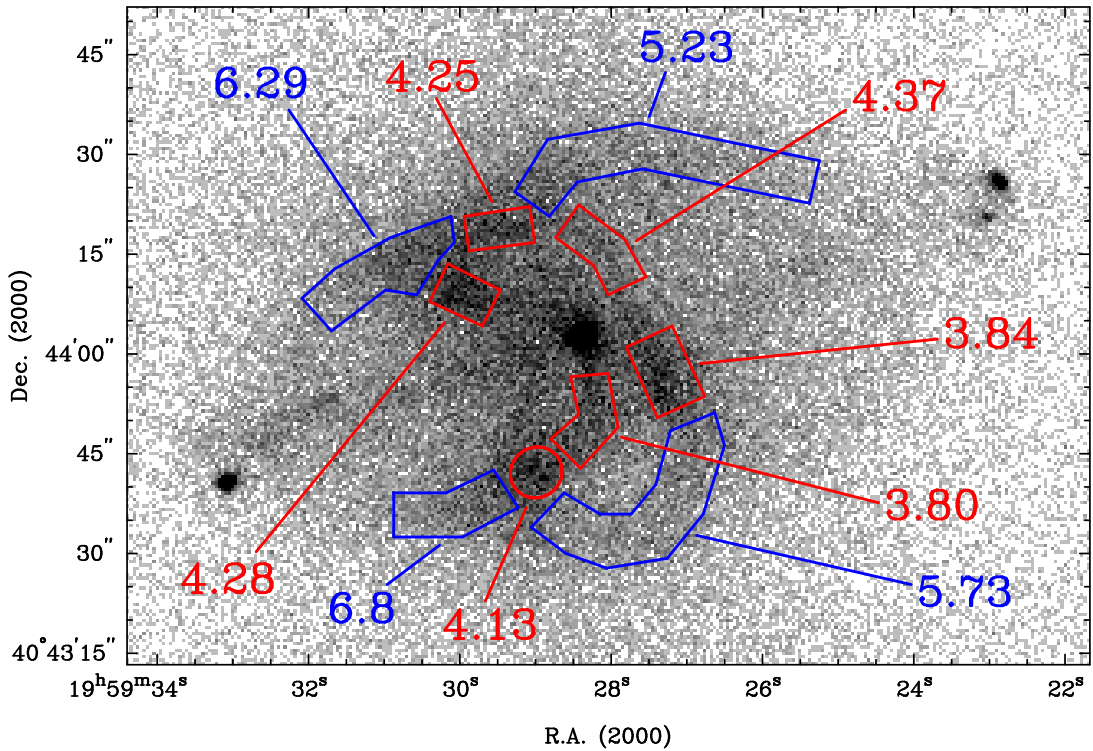}
\caption{An unsmoothed image of the central region of Cyg A in the 0.75 -- 8 keV
band. The areas indicated by solid lines mark regions from which spectra of the
X-ray emission were extracted and modelled with MEKALs; the numbers indicate
the resulting gas temperature in keV. Temperatures of regions marked in blue 
are hotter than the innermost cluster shell (4.60 keV), while those marked in
red are cooler. It is noteable that the hotter regions are associated with the
limb-brightened edges of the cavity, while the cooler regions are in the 
``belts'' which traverse or encircle the cavity along its minor axis, and may
represent the outer regions of the accretion disk.}
\label{Fig. 3}
\end{figure*}

\subsection{Mach Number and Shock Velocity of Expanding Cavity}

We interpret the hot regions elongated along the N and S edges of the cavity 
as intracluster gas shocked by the expanding cavity.
The fact that these hot regions are visible at all implies that the stand-off
distance of the shock is comparable to or larger than Chandra's resolution
(about 1 kpc).
The deprojected temperature
of the innermost shell of the cluster, $T_{\rm 1}$ = 4.60 keV, is taken as the
pre-shock temperature. The average temperature -- $T_{\rm 2}$ = 6.01 keV -- of 
the
four ``blue'' regions on the N and S edges of the cavity is assumed to be
representative of the postshock temperature. 

For a nonradiating shock, equations (10-19) and (10-20) of Spitzer (1978) are

$${\frac{p_{2}}{p_{1}}} = \frac{2\gamma M^{2}}{\gamma + 1} - \frac{\gamma -1}{\gamma + 1}\eqno(1)$$

\noindent
and

$${\frac{u_{2}}{u_{1}}} = {\frac{\rho_{1}}{\rho_{2}}} =
\frac{\gamma -1}{\gamma + 1} + \frac{2}{(\gamma + 1) M^{2}},\eqno(2)$$

\noindent
where $p_{1}$ ($p_{2}$), $u_{1}$ ($u_{2}$) and $\rho_{1}$ ($\rho_{2}$)
represent the pre-shock (post-shock) pressure, velocity and density,
$\gamma$ is the ratio of the specific heats and $M$ the Mach number. 
We have assumed that the magnetic field exterior to the cavity
is sufficiently weak that it does not affect the dynamics.
From
these equations, the ratio of the post-shock ($T_{2}$) to pre-shock 
($T_{1}$) temperatures is

$${\frac{T_{2}}{T_{1}}} = \frac{[2 \gamma M^{2} - (\gamma - 1)][(\gamma - 1) + (2/M^{2})]}{(\gamma + 1)^2}.\eqno(3)$$

\noindent
Equation (3) can be inverted to a quadratic equation in $M^{2}$. Putting
$\gamma$ = 5/3, $T_{2}$ = 6.01 keV, and $T_{1}$ = 4.60 keV, we thus find the
Mach number of the shock driven by the expanding cavity to be $M$ = 1.32. We
note in passing that Alexander \& Pooley (1996) also concluded that the hot
spot advance speed is only mildly supersonic ($M$ $\sim$ 2) on completely
different grounds.

The thermal bremsstrahlung emissivity depends only weakly on temperature
($\epsilon_\nu$ $\propto$ $T^{-1/2}$). We can thus use the jump in emissivity
at the limb-brightened edge of the cavity (a factor of $\simeq$ 3.1 in p.a.
30$^{\circ}$ -- 40$^{\circ}$
from the nucleus) to make a semi-independent estimate of $M$.
We find a density
ratio $\rho_{1}/\rho_{2}$ $\simeq$ 0.67, and a Mach number $M$ $\simeq$ 1.3
(equation 2).
However, within
the uncertainties of both methods, it is possible that Cyg A is simply expanding
at the sound speed of the external intracluster gas. We shall,
nevertheless, continue to assume $M$ = 1.3 in the calculations that follow.
Both methods have their uncertainties from measurements, deprojections and
the simple analytical model, but the excellent agreement between the two is
encouraging.

The shock velocity may now be found from the relation (Spitzer 1978, equation
10-21)

$$u_{1}^{2} = \frac{\gamma k T_{1} M^{2}}{\mu},\eqno(4)$$

\noindent
where $\mu$ is the mean mass per particle and $k$ is the Boltzmann constant.
Assuming the above values for $\gamma$, $T_{1}$ and $M$, and that
$\mu$ = 1.0 $\times$ 10$^{-24}$ gm, we find $u_{1}$ = 1,460 km s$^{-1}$. All
values derived so far are independent of the Hubble constant. Because we have
used gas with a range of azimuthal angles around the cavity to derive this 
shock velocity (at least in the ``temperature method''), we will take
$u_{1}$ = 1,460 km s$^{-1}$ as an approximation
to the angle-averaged velocity of the shock driven by the expanding
cavity.

\subsection{Luminosities, Timescales and Pressures}

The total kinetic power of the expanding cavity is

$$L_{exp} \simeq \frac{1}{2} \rho_{1} u_{1}^{3} A,\eqno(5)$$

\noindent
where $A$ is the total area of the shock surface. We assume this surface has a
prolate spheroidal shape, with major axis in the plane of the sky. From the
X-ray image, we adopt a semi-major axis of a = 62$^{\prime\prime}$ and a 
semi-minor axis of b = 28$^{\prime\prime}$ and find $A$ = 1.7 $\times$
10$^{47}$ cm$^{2}$. From the deprojection of the cluster emission, the density
of the innermost cluster shell, which we take to be the pre-shock density, is
$\rho_{1}$ = 4.4 $\times$ 10$^{-26}$ gm cm$^{-3}$. With
$u_{1}$ = 1,460 km s$^{-1}$, we have 
$L_{exp}$ $\simeq$ 1.2 $\times$ 10$^{46}$ erg s$^{-1}$. For a black hole
mass of (2.5 $\pm$ 0.7) $\times$ 10$^{9}$ M$_{\odot}$ (Tadhunter et al. 2003),
$L_{exp}/L_{Edd}$ is 0.037.

$L_{exp}$ exceeds the total radio emission of Cyg A
($L_{R}$(10 MHz -- 400 GHz) = 9.0 $\times$ 10$^{44}$ erg s$^{-1}$) and the
total cluster gas X-ray emission in the 2 to 10 keV band 
($L_{X}$(2 -- 10 keV) = 1.2 $\times$ 10$^{45}$ erg s$^{-1}$. These values are
from Carilli \& Barthel (1996) adjusted to our assumed luminosity distance.
The dominant current power source in Cyg A thus appears to be the expansion of
the jet-driven cavity which heats the surrounding intracluster gas.
$L_{exp}$ represents a lower limit to the current power of the two jets.

We assume the cavity maintains its axial ratio ($a/b$ $\simeq$ 2.2) as it 
expands. The shock velocity along the minor axis would then be approximately
$u_{b}$ $\simeq$ $(b/a)^{1/2}$$u_{1}$ = 992 km s$^{-1}$ and along the major
axis $u_{a}$ $\simeq$ $(a/b)^{1/2}$$u_{1}$ = 2,150 km s$^{-1}$. Assuming these
velocities
represent the velocities of expansion of the cavity and
have been constant over the lifetime of the radio source, the age
of Cyg A is t$_{\rm age}$ $\simeq$ 3.0 $\times$ 10$^{7}$ yrs.
This value is within a factor of two of
other estimates of the age based on the radio properties
(e.g. Begelman \& Cioffi 1989;
Carilli et al. 1991; Kaiser \& Alexander 1999).
We emphasise that any net infall (``cooling flow'' or gravitational infall)
of the pre-shock gas has been assumed to be zero here, as we have no way of
determining it.

The cooling timescale of the hot, postshock gas, with $\rho_{2}$ $\simeq$
6.5 $\times$ 10$^{-26}$ gm cm$^{-3}$ and
$T_{2}$ $\simeq$ 7.0 $\times$ 10$^{7}$ K, is t$_{\rm cool}$ $\simeq$
2.5 $\times$ 10$^{8}$ yrs. This is a factor of 8 times larger than the age
of Cyg A. Thus, the expanding cavity is currently heating up the surrounding
gas; cooling of the gas is a slower process. This finding that
t$_{\rm cool}$ $>$ t$_{\rm age}$ supports our assumption of a nonradiating
shock.

The lateral expansion of the cavity is driven by its internal pressure
(Scheuer 1974):

$$\rho_{1} u_{b}^{2} = p_{c}\eqno(6)$$

\noindent
With the values given above for $\rho_{1}$ and $u_{b}$, we have
$p_{c}$ $\simeq$ 4.4 $\times$ 10$^{-10}$ dyne cm$^{-2}$. If this pressure
is produced by (tangled) magnetic fields and isotropic relativistic
particles (so $p_{c}$ = $u_{c}/3$) in equipartition, the magnetic field in the
cavity is $B_{c}$ $\simeq$ 1.2 $\times$ 10$^{-4}$ gauss. This value is close to 
the equipartition magnetic field strength in the lobes calculated from their 
radio emission (6 $\times$ 10$^{-5}$ gauss for our distance, Winter et al.
1980; (4.5 -- 6.5) $\times$ 10$^{-5}$ gauss, Carilli et al. 1991;
(5 -- 7) $\times$ 10$^{-5}$ gauss, Alexander \& Pooley 1996). The discrepancy between our value of $B_{c}$ and that obtained from the radio
observations may not be as significant as it seems given the uncertainties in both
estimates.

We note, in addition, that all the equipartition field strengths in the lobes
quoted above assumed equal energy densities in positive particles and electrons
(C. Carilli, P. Alexander, private communications). If it is assumed, instead,
that the
positive particles have 100 times the energy density of the electrons (as is
the case in our Galaxy), then the equipartition field strengths would be 3.7 
times larger and the pressure in cosmic rays and magnetic fields 14 times 
larger. Such a pressure is larger than deduced above from equation (6),
so we do not favor the notion that the energy density in positive particles
is 100 times that in electrons, but we cannot rule it out.

We have found the shock velocity along the major axis of the cavity to be
$u_{a}$ $\simeq$ 2,150 km s$^{-1}$. This agrees well with the hot spot
advance speed of 1,500 km s$^{-1}$ found by Alexander \& Pooley (1996),
assuming ram pressure confinement and that the whole front surface of the
radio source moves forward with the same speed.

\section{Conclusions}

In this paper, we have explored the dynamics of the cavity (or ``cocoon'')
associated with Cygnus A. The primary source of information was Chandra
X-ray imaging and spectroscopy, from which densities and temperatures of the
thermal gas in the intracluster gas around the cavity were obtained.

Our primary assumption is that the hotter (6.0 keV) gas in the limb-brightened
regions at the N and S edges of the cavity has been shocked by the expanding
cavity. From this assumption, we infer the Mach number (1.3) and
velocity (1,460 km s$^{-1}$) of the shock. The same numbers are found from the
density jump.
If, instead, some of this gas has not been shocked, we will have
underestimated the postshock temperature and thus also the Mach number, 
expansion velocity and power of the expanding cavity. The kinetic power of
the expanding cavity
($\simeq$ 1.2 $\times$ 10$^{46}$ erg s$^{-1}$) exceeds the total radio emission
in the 10 MHz to 400 GHz band by a factor of $\simeq$ 13 and the total X-ray
emission from the cluster in the 2 -- 10 keV band by a factor of $\simeq$ 10.
The cooling time of the shocked gas (2.5 $\times$ 10$^{8}$ yrs) is longer than
the age of the radio source by a factor of $\simeq$ 8. It thus appears that
most of the power currently emitted in the relativistic beams which fuel the
radio source goes into heating the intracluster medium. We find that the
pressure inside the cavity (4.4 $\times$ 10$^{-10}$ dynes cm$^{-2}$) exceeds the minimum pressure in magnetic fields and cosmic rays in the radio
lobes. This difference may not be significant, so we find no
conclusive evidence
for any other entity contributing pressure inside the cavity. A further
consequence is that the
ratio of the energy density in positive relativistic particles to that in
relativistic electrons in Cyg A is less than the value of 100 found
in our Galaxy.

If the cavity shock is weak ($M$ = 1.3) in a powerful, young radio galaxy like
Cyg A, then it seems that, in general, heating of the intracluster medium
by radio galaxies
must be through the media of weak shocks and sound waves. Strong shocks are
implausible (cf. Heinz 2003).

The accuracy of our results is limited by the simplified analytical model used.
Further work,
combining the results of the X-ray observations with modern computational
simulations of Cyg A, would be well worthwhile.

\section{Acknowledgements}

This research was partially supported by NASA through grants NAG 81027 and
NAG 513065
to the University of Maryland. We thank Patrick Shopbell for making Fig. 2,
Rick Perley for providing the 6 cm radio map and Eve Ostriker and Mark
Birkinshaw for
a helpful conversation.


\end{document}